\documentclass[aps,prl,twocolumn,tightenlines,showpacs,preprintnumbers,nofootinbib]{revtex4}
\usepackage{amssymb,latexsym}
\usepackage{amsmath,amsbsy}
\usepackage{epsfig,bm}
\DeclareMathOperator{\sign}{\text{sign}}

\begin{document}

\def\a{{\alpha}}
\def\b{{\beta}}
\def\d{{\delta}}
\def\D{{\Delta}}
\def\e{{\varepsilon}}
\def\g{{\gamma}}
\def\G{{\Gamma}}
\def\k{{\kappa}}
\def\l{{\lambda}}
\def\L{{\Lambda}}
\def\m{{\mu}}
\def\n{{\nu}}
\def\o{{\omega}}
\def\O{{\Omega}}
\def\S{{\Sigma}}
\def\s{{\sigma}}
\def\th{{\theta}}
\def\x{{\xi}}
\def\Pperp{{\mathbf{P}^{\perp}}}
\def\Pplus{{P^{+}}}
\def\kperp{{\mathbf{k}^{\perp}}}
\def\kplus{{k^{+}}}
\def\dperp{{\mathbf{\Delta}^\perp}}
\def\kpperp{{\mathbf{k}^{\prime \perp}}}
\def\xpp{{x^{\prime \prime}}}
\def\xp{{x^\prime}}
\def\xt{{\tilde{x}}}
\def\eps{{\varepsilon}}
\def\kppperp{{\mathbf{k}^{\prime \prime \perp}}}
\def\Pp{{P^\prime}}
\def\lp{{\lambda^\prime}}

\def\ol#1{{\overline{#1}}}

\preprint{NT-UW 04-09}
\title{Double distributions: Loose ends}
\author{B.~C.~Tiburzi}
\affiliation{Department of Physics,  
	University of Washington,     
	Box 351560,
	Seattle, WA 98195-1560}
\date{\today}

\begin{abstract}
We point out that double distributions need not vanish at their boundary. Boundary 
terms do not change the ambiguity inherent in defining double distributions; instead, boundary
conditions must be satisfied in order to switch between different decompositions. 
We analyze both the spin zero and spin one-half cases.
\end{abstract}


\maketitle

QCD factorization provides the way to access information experimentally 
about the non-perturbative quark and gluon 
substructure of hadrons. In recent years, much attention has been generated 
by hard exclusive reactions such as deeply virtual Compton scattering, and the hard
electroproduction of mesons~\cite{Muller:1994fv}. 
The non-perturbative structure functions entering these reactions are generalized parton distributions (GPDs),
see the reviews~\cite{Ji:1998pc}.

The polynomiality property required of the Mellin moments of GPDs is elegantly explained by the formalism 
of double distributions (DDs)~\cite{Radyushkin:1997ki,Polyakov:1999gs,Belitsky:2000vk}.
Phenomenological modeling of GPDs is almost exclusively done utilizing parametrized DDs. 
Experimentally one cannot access the DDs directly, only indirectly through the $H(x,\x,t)$ and $E(x,\x,t)$ GPDs. 
In Ref.~\cite{Teryaev:2001qm}, the ambiguity inherent 
in defining DDs for the pion was likened to the gauge ambiguity of the vector potential of a two-dimensional magnetic field.
Here we extend this analysis to the proton case.  But first, we review the pion case and tie up the loose ends 
relating to the non-vanishing value of the DDs at their boundary. Additionally we compare various forms of DDs.

For the pion, we have the following decomposition of the non-diagonal twist-two matrix elements
in terms of various form factors $A_{nk}(t)$ and $B_{nk}(t)$ 
\begin{multline} 
\langle P^\prime| 
\ol \psi (0) \gamma^{\{\mu}i\tensor D {}^{\mu_1} \cdots i\tensor D {}^{\mu_n\}} \psi(0)
| P\rangle  
\\ 
=  \sum_{k=0}^{n} \frac{n!}{ k! (n-k)!}
\left [ 2 \ol P {}^{\{\mu} A_{nk}(t) 
- \D^{ \{ \mu}  B_{nk}(t)  \right] 
\phantom{spac} \\
\times
\ol P {}^{\mu_1} \cdots \ol P {}^{\mu_{n-k}} 
\left( - \frac{\D}{2}\right)^{\mu_{n-k+1}} \cdots \left( - \frac{\D}{2}\right)^{\mu_{n}\}} \label{eqn:pionmoments}
,\end{multline}
where the gauge covariant derivative $\tensor D = (\overset{\rightarrow} D -  \overset{\leftarrow}D)/2$, 
$\{ \ldots \}$ denotes the symmetrization and trace subtraction performed on Lorentz indices, $\ol P = (P' + P)/2$, $\D = P' - P$, 
and $t = \D^2$. Often we treat the $t$-dependence as implicit below.
Above, $T$-invariance restricts $A_{nk}(t) = 0$ for $k$ odd, and $B_{nk}(t) = 0$ for $k$ even. 
There is manifest arbitrarity in the twist-two form factors appearing in Eq.~\eqref{eqn:pionmoments}.
The particular decomposition above can be used to define two DDs for the pion. 
These DDs are generating functions for the twist-two form factors
\begin{equation} \label{eqn:generate2}
\begin{pmatrix}
A_{nk} \\
B_{nk}
\end{pmatrix}
= \int_{\b,\a}
\b^{n - k} \a^k 
\begin{pmatrix}
F(\b,\a) \\
G(\b,\a) 
\end{pmatrix}
.\end{equation} 
Above we have abbreviated the integration as $\int_{\b,\a} = \int_{-1}^{1} d\b \int_{-1 + |\b|}^{1 - |\b|} d\a$. 
As a consequence of $T$-invariance, the function $F(\b,\a)$ is even in $\a$, while
$G(\b,\a)$ is odd.

Summing up the moments in Eq.~\eqref{eqn:pionmoments}, 
these DD functions then appear in matrix elements of the light-like separated quark bilinear operator
$\mathcal{M}(\ol  P \cdot z, \D \cdot z) \equiv  \langle P^\prime| 
\ol \psi \left( - z/2 \right) \rlap \slash z \psi \left( z/2 \right) | P \rangle$, 
\begin{align} 
\mathcal{M}( \ol P \cdot z, \D \cdot z) 
& = \int_{\b,\a}  e^{ - i \b \ol P \cdot z + i \a \D \cdot z / 2}
\notag
\\ 
& \times
\left[ 2 \ol P \cdot z   \, F(\b,\a)
- 
\D \cdot z \,  G(\b,\a)
\right] \label{eqn:bilocal2}
,\end{align}
where $z^2 = 0$. 
Now we define the pion GPD 
\begin{equation} \label{eqn:pionGPD}
H(x,\x) = \frac{1}{2} \int \frac{dz^-}{2\pi } 
e^{i x \ol P {}^+ z^-} \mathcal{M}( \ol P \cdot z, \D \cdot z)
,\end{equation}
with the usual definition $\x = - \D^+ / 2 \ol P {}^+$.

Any other choice of generating functions for the moments must lead to the same GPD. 
To expose this ambiguity, we follow \cite{Teryaev:2001qm}. 
Integration of Eq.~\eqref{eqn:bilocal2} by parts produces surface terms that in general do not vanish. 
We are careful about this point because previous models \cite{Radyushkin:1997ki,Tiburzi:2002tq} 
indeed have non-vanishing contributions on the boundary $|\b| = 1 - |\a|$. 
Mathematically the DDs must vanish only at the corners of support: $\delta(\a) \delta(|\b| - 1)$, $\delta(\b) \delta(|\a| -1)$,
else the form factors in Eq.~\eqref{eqn:generate2} do not fall off in the space of moments. Physically the DDs' vanishing at the first set of 
corners is tied via Eq.~\eqref{eqn:pionGPD} to the vanishing of the GPDs at $x = \pm 1$, 
which is known from perturbative QCD~\cite{Yuan:2003fs}. Vanishing of the DDs 
at the second set of corners implies the continuity of GPDs at the crossover ($x = \x$), which in turn ensures factorization.
Boundary contributions are thus not ruled out and the DDs appear to be loose at the ends.
Phenomenological parameterizations of DDs must generally include such terms. 
The complete result of integrating by parts can be expressed in the form
\begin{eqnarray}
\mathcal{M} ( \ol P \cdot z, \D \cdot z)  
&=& - 2 i 
\int_{\b,\a} 
N(\b,\a)
\, e^{ - i \b \ol P \cdot z + i \a \D \cdot z / 2} 
\notag  \\ 
&& + 8 i \int_0^1 d\a \, \cos ( \a \D \cdot z / 2 )  
\notag \\
&&  \times \Big\{
\cos \left[ (1-\a) \ol P \cdot z \right]
S^+(\a) \notag \\
&&  - i \sin \left[(1-\a)\ol P \cdot z \right]
S^-(\a)
\Big\} \label{eqn:surface}
,\end{eqnarray}
where $N(\b,\a) = \frac{\partial}{\partial\b} F(\b,\a)+ \frac{\partial}{\partial\a} G(\b,\a)$ and
$S^\pm(\a) =  F^\mp(1-\a,\a) + G^\pm(1-\a,\a)$. 
In deriving Eq.~\eqref{eqn:surface}, we made use of the $\a$-symmetry of DDs
and additionally have used $\pm$-distributions ($F^\pm$ and $G^\pm$), 
which are even and odd functions of $\b$, respectively.

Consider an arbitrary potential function $\chi(\b,\a)$ which is odd with respect to $\a$. 
We can decompose $\chi$ in terms of its even and odd parts with respect to $\b$, 
namely $\chi^\pm(\b,\a)$ where
$\chi^\pm(-\b,\a) =  \pm \chi^\pm(\b,\a)$.
The potential function can be used to generate the DD transformation
\begin{equation}
\begin{pmatrix}
F^\pm (\b,\a) \\
G^\pm (\b,\a)
\end{pmatrix}
\to  
\begin{pmatrix} 
F^\pm(\b,\a) + \frac{\partial}{\partial \a} \chi^\pm(\b,\a) \\
G^\pm(\b,\a) - \frac{\partial}{\partial \b} \chi^\mp(\b,\a)
\end{pmatrix}
\label{eqn:Ftransform} 
.\end{equation}
Notice this transformation preserves both the $\a$- and $\b$-symmetry of the $\pm$-distributions.
Next we observe Eq.~\eqref{eqn:surface} is invariant under this transformation provided
the potential also satisfies two boundary conditions
\begin{equation} \label{eqn:boundary}
\frac{\partial}{\partial \a} 
\chi^\pm(\b,\a) 
\Bigg|_{\b = 1 - \a}
= 
\frac{\partial}{\partial \b}
\chi^\pm(\b,\a) 
\Bigg|_{\b = 1 - \a}
.\end{equation}
Above both $\a$ and $\b$ are positive and the corresponding boundary conditions 
for the full range of DD variables can be found trivially due to 
the symmetry properties. 
Without the boundary conditions Eq.~\eqref{eqn:boundary}
on the potential function, the transformation
generated by Eq.~\eqref{eqn:Ftransform} is that of~\cite{Teryaev:2001qm}.

To convert the DDs in Eq.~\eqref{eqn:bilocal2} to the Polyakov-Weiss ``gauge'' \cite{Polyakov:1999gs}, 
in which there is only an $F$-type double distribution and $D$-term, we use the potential specified by
\begin{multline}
\chi_o^\pm(\b,\a)  = - \frac{1}{2}
\Bigg[
\int_{-1 + |\a|}^\b d\b' \, G^\mp(\b',\a) 
\\
- \int_\b^{1 - |\a|} d \b'  \, G^\mp(\b',\a)
-  \frac{1}{2} \left( 1 \mp 1 \right) \sign (\b) D(\a)
\Bigg]  \label{eqn:PWpot}
,\end{multline} 
where $D(\a)$ is the $D$-term given by
\begin{equation}
D(\a) = \int_{- 1+ |\a|}^{1 - |\a|} d\b \, G^+(\b,\a)
.\end{equation}
One can verify that the potential specified by 
Eq.~\eqref{eqn:PWpot} satisfies the boundary conditions Eq.~\eqref{eqn:boundary} provided  
one also assumes the GPD is continuous, i.e., $D(\pm 1;t) = 0$.
Under the transformation generated by Eq.~\eqref{eqn:PWpot}, 
the bilocal matrix element reads
\begin{multline} \label{eqn:bilocal3}
\mathcal{M} (\ol P \cdot z, \D \cdot z)
= 
\int_{\b,\a} e^{ - i \b \ol P \cdot z + i \a \D \cdot z / 2}
\\
\times
\left[ 2 \ol P \cdot z \, \,  F_o(\b,\a) 
- 
\D \cdot z \, \, \delta(\b) \, D(\a)
\right]  
,\end{multline}
where the resulting DD  $F_o(\b,\a;t)$ is given by
\begin{equation} \label{eqn:PWF}
F_o(\b,\a) = F(\b,\a) + \frac{\partial}{\partial \a} \left[ \chi^+_o(\b,\a) + \chi^-_o(\b,\a) \right] 
.\end{equation}

A perhaps more interesting choice of gauge is what we call the Drell-Yan gauge. 
It is specified by the potential function
\begin{multline}
\chi_1^\pm(\b,\a)  = - \frac{1}{2}
\Bigg[
\int_{-1 + |\b|}^\a d\a' \, F^\pm(\b,\a') \\
- \int_\a^{1 - |\b|} d \a'  \, F^\pm(\b,\a')
-  \sign (\a) \ol D {}^\pm(\b)
\Bigg],  
\label{eqn:DYpot}
\end{multline} 
where the $\ol D$-term $\ol D(\b)$ 
is given by $\ol D (\b) = \ol D {}^+ (\b) + \ol D {}^- (\b)$, with
\begin{equation}
\ol D {}^\pm (\b) = \int_{- 1+ |\b|}^{1 - |\b|} d\a \, F^\pm (\b,\a)
.\end{equation}
Again one can verify that the potential specified by 
Eq.~\eqref{eqn:DYpot} satisfies the boundary conditions 
Eq.~\eqref{eqn:boundary}.
Under this transformation, the bilocal matrix element reads
\begin{multline} \label{eqn:bilocal4}
\mathcal{M}(\ol P \cdot z, \D \cdot z)
= 
\int_{\b,\a} e^{ - i \b \ol P \cdot z + i \a \D \cdot z / 2}
\\
\times
\left[ 2 \ol P \cdot z   \, \, \delta(\a) \,  \ol D (\b)  
- 
\D \cdot z  \, \, G_1(\b,\a) 
\right] 
,\end{multline}
where the resulting DD $G_1(\b,\a)$ is given by
\begin{equation}
G_1(\b,\a) = G(\b,\a) - \frac{\partial}{\partial \b} 
\left[ 
\chi^+_1(\b,\a) + \chi^-_1(\b,\a) 
\right]
.\end{equation}

The Drell-Yan gauge is particularly interesting from the perspective of GPDs. 
Inserting Eq.~\eqref{eqn:bilocal4} into the definition of the pion GPD 
Eq.~\eqref{eqn:pionGPD}, we have (now reinstating the $t$-dependence)
\begin{equation} \label{eqn:nice}
H(x,\x, t) = \ol D(x;t)  
+ \x \int_{\b,\a} \delta( x - \b - \x \a) \, G_1(\b,\a;t)  
.\end{equation}
Notice the contribution to the GPD from the $\ol D$-term is independent of $\x$ and the reduction relations are simply
\begin{equation}
f_1(x) = H(x,0,0) = \ol D (x, 0)
,\end{equation}
for the quark distribution and
\begin{equation}
F(t) = \int dx \, H(x, \x, t)  = \int dx \, \ol D (x;t)
,\end{equation}
for the pion form factor. Thus the $\ol D$-term has a simple physical interpretation as the $x$-integrand of the
form factor calculated in the Drell-Yan frame ($\x = 0$).  Changing to a frame where $\x \neq 0$, we 
have Eq.~\eqref{eqn:nice}. Note, the second term in the GPD proportional to $\x$ contributes nothing to the form factor
because of Lorentz invariance. This is maintained by the $\a$-symmetry of the DD. The GPD, by contrast, is manifestly affected 
by the change in frame. At fixed $t$, the change to $\x \neq  0$ requires a dynamical light-front rotation.
Despite claims \cite{Mukherjee:2002gb}, the $\ol D$-term is all that one can learn about DDs from the Drell-Yan expression
for the form factor \cite{Tiburzi:2002kr}.

To describe the pion GPD, we have resorted to using two DDs $F$ and $G$ because 
these are encountered in actual calculations \cite{Tiburzi:2002tq}. The GPD, however, 
can be viewed as the projection of a single DD function \cite{Belitsky:2000vk}, see 
also \cite{Belitsky:2000vx}. 
To see this representation one must use a non-trivial transformation generated by 
$\chi_f(\b,\a)$, where
\begin{equation}
\chi^\pm_f(\b,\a) = \chi^\pm_0(\b,\a) + \a \int_{-1 + |\a|}^{1 -|\a|} d\b' W(\b,\b') f^\mp(\b',\a)
,\end{equation}
with $W(\b,\b') = \theta(\b) \theta(\b' -\b)  - \theta(-\b) \theta(\b -\b')$.
The new DD $f(\b,\a)$, which is even in $\a$, is implicitly defined by the result of the 
transformation 
\begin{multline} \label{eqn:bilocal5}
\mathcal{M}(\ol P \cdot z, \D \cdot z)
= 
\int_{\b,\a} e^{ - i \b \ol P \cdot z + i \a \D \cdot z / 2}
\\
\times \left[ 2 \ol P \cdot z \, \b     
- 
\D \cdot z  \,  \a
\right] f(\b,\a)
.\end{multline}
The boundary conditions are met provided
\begin{equation}
f^\mp (1 - \a,\a) = F_o^\pm (1 - \a, \a),  
\end{equation}
where $F_o$ is given in Eq.~\eqref{eqn:PWF}.

Now we address the ambiguities of proton DDs.
The non-diagonal proton matrix elements of twist-two operators can be decomposed 
into form factors $A_{nk}(t)$, $B_{nk}(t)$ and $C_{nk}(t)$
\begin{multline} 
\langle P^\prime,\lp | 
\ol \psi(0) \gamma^{\{\mu}i\tensor D {}^{\mu_1} \cdots i\tensor D {}^{\mu_n\}} \psi(0)
| P,\l \rangle  
\\
= \ol u_{\lp}(\Pp)
\sum_{k=0}^{n} 
\frac{n!}{ k! (n-k)!}
\Bigg[ 
\gamma^{\{ \mu }  A_{nk}(t) 
\phantom{spacerererer}
\\
+ \frac{i \sigma^{ \{ \mu \nu } \D_\nu}{2 M} B_{nk}(t) 
- \frac{\D^{ \{ \mu} }{4 M} C_{nk}(t)
\Bigg] u_{\l}(P)
\phantom{spacer}  
\\
\times
\ol P {}^{\mu_1} \cdots \ol P {}^{\mu_{n-k}} 
\left( - \frac{\D}{2}\right)^{\mu_{n-k+1}} \cdots \left( - \frac{\D}{2}\right)^{\mu_{n}\}}. 
\label{eqn:moments}
\end{multline}
$T$-invariance forces $A_{nk}(t) = B_{nk}(t) = 0$ for $k$ odd, and 
$C_{nk}(t) = 0$ for $k$ even. 
There are three Dirac structures in the above decomposition since in general the 
twist-two currents are not conserved.

The above decomposition can 
be used to define three double distributions as generating functions for the twist-two form factors
\begin{equation} \label{eqn:generate}
\begin{pmatrix}
A_{nk} \\
B_{nk} \\
C_{nk}
\end{pmatrix}
= 
\int_{\b,\a}
\b^{n - k} \a^k 
\begin{pmatrix}
F(\b,\a) \\
K(\b,\a) \\
G(\b,\a)
\end{pmatrix}
.\end{equation} 
$T$-invariance implies the functions $F(\b,\a)$ and $K(\b,\a)$ are even in $\a$, while
$G(\b,\a)$ is odd. Eq.~\eqref{eqn:moments} represents a ``physical'' gauge for proton DDs since these are 
encountered in actual calculations \cite{Tiburzi}.

Summing up the moments in Eq.~\eqref{eqn:moments}, 
the DD functions then appear in matrix elements of the light-like separated quark bilinear operator
$\mathcal{M}^{\lp,\l}(\ol P \cdot z, \D \cdot z)  \equiv 
\langle P^\prime,\lp | \ol \psi \left( - z/2 \right) \rlap\slash z \psi \left( z/2 \right)  | P,\l \rangle$, 
\begin{multline} \label{eqn:bilocal}
\mathcal{M}^{\lp,\l} (\ol P \cdot z, \D \cdot z) 
= 
\int_{\b,\a}
e^{ - i \b \ol P \cdot z + i \a \D \cdot z / 2}
\ol u_{\lp}(\Pp) 
\\
\times \Bigg[ \rlap \slash z  F(\b,\a)
+
\frac{i \sigma^{\mu \nu} z_\mu \D_{\nu}}{2 M}
K(\b,\a) 
-
\frac{\D \cdot z}{4 M} G(\b,\a)
\Bigg] u_{\l}(P) 
\end{multline}

Now we define the light-cone correlation function 
\begin{equation} \label{eqn:lcc}
\mathcal{M}^{\lp,\l}(x,\x) = \frac{1}{2} \int \frac{dz^-}{2\pi} 
e^{i x \ol P {}^+ z^-} \mathcal{M}^{\lp,\l}(\ol P \cdot z, \D \cdot z) 
.\end{equation}
This correlation function can be written in terms of the two independent GPDs $H(x,\x)$ and $E(x,\x)$ 
\begin{multline} \label{eqn:lccor}
\mathcal{M}^{\lp,\l}(x,\x) = 
\frac{1}{2 \ol P {}^+} \ol u_{\lp}(\Pp) 
\\
\times \left[ 
\gamma^+ H(x,\x)
+ 
\frac{i \sigma^{+\nu} \D_\nu }{2 M}
E(x,\x)     
\right] u_{\l}(P)   
.\end{multline}
Inserting the DD decomposition Eq.~\eqref{eqn:bilocal} into the correlator in Eq.~\eqref{eqn:lcc}, we 
can express the GPDs as projections of the DDs 
\begin{equation}
\begin{pmatrix}
H(x,\x) \\
E(x,\x)
\end{pmatrix}
= \int_{\b,\a}
\delta(x - \b - \x \a) 
\begin{pmatrix}
F(\b,\a) + \x G(\b,\a) \\
K(\b,\a) + \x G(\b,\a)
\end{pmatrix}
,\end{equation}
from which we can view the $\x$-dependence of GPDs as arising from different slices of Lorentz invariant DDs. 
Due to the symmetry of the DDs with respect to $\a$, the GPDs $H(x,\x,t)$ and $E(x,\x,t)$ 
are both even functions of the skewness parameter $\x$.

Utilizing the Gordon identities,  
we can rewrite the bilocal matrix element in Eq.~\eqref{eqn:bilocal} as 
\begin{multline} \label{eqn:curl}
\mathcal{M}^{\lp,\l}(\ol P \cdot z, \D \cdot z)
= \frac{ \ol u_{\lp}(\Pp) }{2 M - \frac{t}{2 M} }
\int_{\b,\a}
e^{ - i \b \ol P \cdot z + i \a \D \cdot z / 2}
\\
\times
\Bigg[  2 \ol P \cdot z   \, G_E(\b, \a)
-  \D \cdot z \,  \tilde{G}(\b,\a)
\\
- i  \varepsilon^{\mu \nu \a \b} z_\mu \D_\nu \ol P_\a \gamma_\b \gamma_5 G_M(\b,\a)
\Bigg]  u_{\l}(P) 
,\end{multline}
where we have defined new double distributions
\begin{align}
G_E(\b,\a) &= F(\b,\a) + \frac{t}{4 M^2} K(\b,\a) \\
G_M(\b,\a) &= F(\b,\a) + K(\b,\a) \\
\tilde{G}(\b,\a) &= \frac{1}{2} \left( 1 - t /4 M^2 \right) G(\b,\a)
\end{align}
in analogy with the Sachs electric and magnetic form factors. 
After integrating by parts, we have
\begin{multline}
\mathcal{M}^{\lp,\l}(\ol P \cdot z, \D \cdot z)
=  - \frac{ i \, \ol u_{\lp}(\Pp)}{2 M - \frac{t}{2 M} }
\Bigg( \int_{\b,\a}
e^{ - i \b \ol P \cdot z + i \a \D \cdot z / 2}
\\
\times
\Bigg[ 
2 N(\b, \a) 
+   
\varepsilon^{\mu \nu \a \b} z_\mu \D_\nu \ol P_\a \gamma_\b \gamma_5 \, G_M(\b,\a)
\Bigg]
\\
\phantom{sp} - 8 \int_0^1 d\a \, \cos ( \a \D \cdot z / 2 ) 
\Big\{
\cos \left[ (1-\a) \ol P \cdot z \right]
S^+(\a)  \\
\phantom{space} - i \sin \left[(1-\a)\ol P \cdot z \right]
S^-(\a) \Big\} \Bigg)   u_{\l}(P), 
\label{eqn:curl2} \end{multline}
where $N(\b,\a) =  \frac{\partial}{\partial \b} G_E(\b,\a) + \frac{\partial}{\partial \a} \tilde{G}(\b,\a)$,
and the boundary terms $S^\pm(\a) = G_E^\mp(1-\a,\a) + \tilde{G}^\pm(1-\a,\a)$.

Now we define a potential function $\chi(\b,\a)$ as above.
The expression Eq.~\eqref{eqn:curl2}  is invariant under the transformation
\begin{equation}
\begin{pmatrix}
G^\pm_E(\b,\a) \\
G^\pm_M(\b,\a) \\
\tilde{G}^\pm(\b,\a)
\end{pmatrix}
\to 
\begin{pmatrix}
G^\pm_E(\b,\a) + \frac{\partial}{\partial \a} \, \chi^\pm(\b,\a) \\ 
G^\pm_M(\b,\a) \\
\tilde{G}^\pm(\b,\a) -  \frac{\partial}{\partial \b} \, \chi^\mp(\b,\a)
\end{pmatrix}
\label{eqn:transf} 
,\end{equation}
provided the boundary conditions on $\chi^\pm$ Eq.~\eqref{eqn:boundary} are satisfied.
Translating the transformation Eq.~\eqref{eqn:transf} 
into the original DDs, we have invariance under
\begin{equation} 
\begin{pmatrix}
F(\b,\a) \\ 
K(\b,\a) \\
G(\b,\a)
\end{pmatrix}
\to 
\begin{pmatrix}
F(\b,\a) +   \frac{\partial}{\partial \a}  \,\chi(\b,\a) \\
K(\b,\a) -   \frac{\partial}{\partial \a}  \,\chi(\b,\a) \\
G(\b,\a) -   \frac{\partial}{\partial \b}  \,\chi(\b,\a) 
\end{pmatrix}
.\label{eqn:trans} 
\end{equation}

As with the pion case, one can convert the proton DDs to Polyakov-Weiss gauge or Drell-Yan gauge
using the potentials $\chi_o^\pm(\b,\a)$ and $\chi_1^\pm(\b,\a)$, respectively.
Additionally two independent 
DDs for the proton can be unmasked using the transformation generated by $\chi_f^\pm(\b,\a)$. 
Each of these transformations for the proton case is understood 
under the replacement $\{F(\b,\a) \to G_E(\b,\a)\}$ and $\{G(\b,\a) \to \tilde{G}(\b,\a)\}$.
In Polyakov-Weiss gauge we recover the result of \cite{Polyakov:1999gs} for the proton. In 
Drell-Yan gauge, we generate an additive $\x$-independent contribution to the combination of GPDs  
$H(x,\x) + \frac{t}{4 M^2} E(x,\x)$; 
this is the $x$-integrand of the Sachs electric form factor in the Drell-Yan frame.
The $\x$ dependence is contained in a term analogous to the one in Eq.~\eqref{eqn:nice}
and characterizes the dynamical light-front rotation to $\x \neq 0$.
In the minimal gauge generated by $\chi_f^\pm(\b,\a)$, 
we see there are only two underlying independent DDs.

There are two additional Drell-Yan type gauges  for the proton. 
Using $\chi_1^\pm(\b,\a)$ as appears in Eq.~\eqref{eqn:DYpot}, 
we generate an additive $\x$-independent contribution to $H(x,\x)$. This 
is the $x$-integrand of the Dirac form factor calculated in the Drell-Yan frame.
Alternately we can use Eq.~\eqref{eqn:DYpot} under the replacement $\{F(\b,\a) \to - K(\b,\a)\}$ to 
generate the $\x$-independent, $x$-integrand of the Pauli form factor in the Drell-Yan frame, 
which contributes to $E(x,\x)$. With Eq.~\eqref{eqn:trans}, we cannot simultaneously find 
Drell-Yan contributions to GPDs from both Dirac and Pauli form factors. 
This limitation arises from Eq.~\eqref{eqn:transf}, as we cannot generate an additive 
contribution to $H(x,\x) + E(x,\x)$ from the $x$-integrand of the Sachs magnetic form factor.
Thus Eq.~\eqref{eqn:curl2} may contain additional freedom.

Above we have seen that DDs do not necessarily vanish at their boundary
and have investigated the consequences of boundary terms on the ambiguity 
inherent to DDs. The potential function that generates a transformation
of DDs must satisfy boundary conditions. We have carried out this 
analysis for both the spin zero and spin one-half cases. Inclusion 
of boundary terms into model DDs is necessary for general phenomenological
parameterizations of GPDs.

\begin{acknowledgments}
We thank W.~Detmold and G.~A.~Miller for useful comments. 
This work was funded by the U.~S.~Department of Energy, grant: DE-FG$03-97$ER$41014$.  
\end{acknowledgments}

\end{document}